\documentclass[article, prl, nobibnotes, secnumarabic, amssymb, superscriptaddress, twocolumn, aps,longbibliography]{revtex4-2}
\usepackage{chemformula}
\usepackage{upgreek}
\usepackage{bm}
\usepackage{mathrsfs}
\usepackage{multirow}
\usepackage{longtable}
\usepackage{braket}
\usepackage{threeparttable}
\usepackage{threeparttablex}
\usepackage{booktabs}
\usepackage{pdfcomment}
\usepackage[normalem]{ulem}
\usepackage{hyperref}
\usepackage{xcolor}
\hypersetup{colorlinks,breaklinks,
            urlcolor=[rgb]{0,0,0.64},
            linkcolor=[rgb]{0,0,0.64},
            citecolor=[rgb]{0,0,0.64},
            filecolor=[rgb]{0,0,0.64}}
\newcommand{\StanfordP}{Department of Physics, Stanford University, Stanford, CA 94305, USA}

\newcommand{\StanfordAP}{Department of Applied Physics, Stanford University, Stanford, CA 94305, USA}
\newcommand{\SIMES}{Stanford Institute for Materials and Energy Sciences, SLAC National Accelerator Laboratory, Menlo Park, CA 94025, USA}
\newcommand{\LASQC}{School of Physics and Astronomy, Shanghai Jiao Tong University, Shanghai 200240, China}
\newcommand{\TDL}{Tsung-Dao Lee Institute, Shanghai Jiao Tong University, Shanghai 201210, China}
\newcommand{\SLAC}{Linac Coherent Light Source, SLAC National Accelerator Laboratory, Menlo Park, CA 94025, USA}

\begin{document}

\title{{Selective strain tuning of nonequilibrium multiferroic dynamics in BiFeO$_\text3$}}
\author{Chenhang~Xu}
\thanks{These authors contributed equally to this work: C.X. and P.L.}
\affiliation{\StanfordAP}
\affiliation{\SIMES}
\author{Patrick~Liu}
\thanks{These authors contributed equally to this work: C.X. and P.L.}
\affiliation{\StanfordAP}
\author{Bo~Zhang}
\affiliation{\LASQC}
\affiliation{\TDL}
\author{Minyong~Han}
\affiliation{\StanfordAP}
\affiliation{\SIMES}
\author{Henry~G.~Bell}
\affiliation{\StanfordAP}
\affiliation{\SIMES}
\author{Cameron~J.~R.~Duncan}
\affiliation{\SLAC}
\author{Yusong~Liu}
\affiliation{\SLAC}
\author{Patrick~L.~Kramer}
\affiliation{\SLAC}
\author{Randy~Lemons}
\affiliation{\SLAC}
\author{Jake~D.~Koralek}
\affiliation{\SLAC}
\author{Alexander~H.~Reid}
\affiliation{\SLAC}
\author{Dong~Qian}
\affiliation{\LASQC}
\affiliation{\TDL}
\author{Harold~Y.~Hwang}
\affiliation{\StanfordAP}
\affiliation{\SIMES}
\author{Alfred~Zong}
\email[Correspondence to: ]{alfredz@stanford.edu}
\affiliation{\StanfordAP}
\affiliation{\SIMES}
\affiliation{\StanfordP}
\date{\today}

\begin{abstract}
Quantum materials are characterized by intertwined orders, and a fundamental goal in condensed matter physics is to achieve their selective control in such a way that one order parameter can be tuned while leaving others largely unaffected. Although significant progress has been made in thermal equilibrium, realizing such control out of equilibrium remains highly challenging. Here we employed \textit{in situ} tensile strain to selectively manipulate the photoinduced lattice dynamics associated with ferroelectric and magnetic orders in multiferroic BiFeO$_3$. By applying MeV ultrafast electron diffraction to freestanding BiFeO$_3$ membranes under tunable strain, we showed that tensile strain markedly suppresses the ultrafast photoinduced reduction of the ferroelectric displacement. By contrast, the photoinduced dynamics of the antiferrodistortive rotation of the oxygen octahedra, which modulates the magnetic order, remains insensitive to strain. Not only do these findings reveal distinct microscopic pathways underlying nonequilibrium multiferroic dynamics, they also establish tunable strain as an effective route for engineering ultrafast phase control in correlated materials.
\end{abstract}

\maketitle

Quantum materials host intertwined orders and complex phase diagrams arising from coupled degrees of freedom~\cite{keimer2015from}. Riding on the rapid progress in ultrafast laser technologies, tailored photoexcitation has been shown to be an effective way to transiently reshape the balance among coexisting orders~\cite{basov2017towards,xu2025timedomain_npj}, as exemplified by the photoinduced competition between superconductivity and charge density wave (CDW) order in cuprates~\cite{wandel2022enhanced} or the competing CDWs in rare-earth tritellurides~\cite{Kogar2020light_NP,zong2021role,Zhou2021Natlate4}. However, the ensuing dynamics are often set by the intrinsic interactions between the coupled orders in equilibrium, hence limiting the on-demand design of nonequilibrium pathways. For instance, in the case of the tritellurides~\cite{zong2021role}, the photoinduced CDW along the \textit{a}-axis is always accompanied by a suppression of the equilibrium CDW along the \textit{c}-axis, making it difficult to steer one response without triggering the other. This limitation highlights the need for additional control parameters to achieve more precise and versatile manipulation of nonequilibrium states of matter.

\begin{figure*}[!t]
    \centering
    \includegraphics[width=1\linewidth]{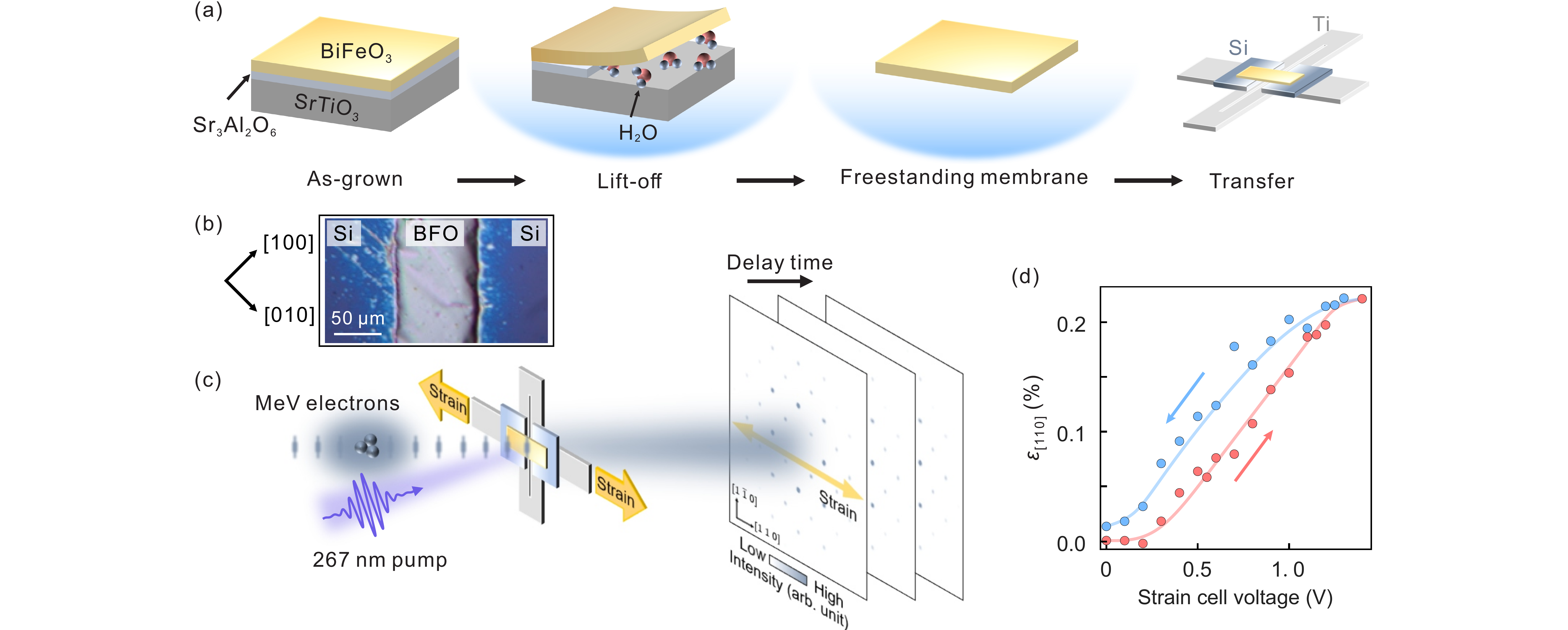}
    \caption{Experimental setup for \textit{in situ} straining of large-area, freestanding thin films in ultrafast electron diffraction.
    (a)~Steps for thin film fabrication: (i)~an as-grown heterostructure is prepared by pulsed laser deposition; (ii)~the sacrificial Sr$_3$Al$_2$O$_6$ layer is selectively etched in water; (iii)~a millimeter-sized freestanding BiFeO$_3$ film floats to the water surface; (iv)~a scoop is used to transfer the film onto the gap in a silicon chip, which is affixed to a titanium plate mounted on a piezoelectric strain actuator (not drawn). 
    (b)~Optical image of the freestanding membrane suspended across the silicon gap.
    (c)~Schematic of the ultrafast electron diffraction setup with \textit{in situ} strain control, where two piezoelectric stacks pull the two sides of the titanium plate to apply strain to the freestanding BiFeO$_3$ membrane.
    (d)~Tensile strain along the [110] direction of BiFeO$_3$ as a function of the voltage applied to the piezoelectric actuator. The strain was calculated based on the momentum space separation of Bragg peaks, whose positions were accurately determined by fitting each peak to a Gaussian function. Arrows indicate the measurement sequence, starting at 0~V.}
    \label{fig:exp_setup}
\end{figure*}

Among the available tuning knobs, strain stands out as a particularly powerful method to access novel states in thermal equilibrium~\cite{Hicks2014Sr2RuO4Science,hong2020extreme,hemme2023tuning,singh024ErTe3_strain_SA,li2025classical}, stabilizing phases such as superconductivity in ruthenium oxides and nickelates~\cite{Uchida2020PRL_RuO2strain,ruf2021strainSC_NCRuO2,ko2025Natursignatures} and ferroelectricity in quantum paraelectrics~\cite{haeni2004roomtemp,fechner2024quance_NMSTO,li2025classical}. Strain can modify lattice parameters, symmetries, and phonon spectra, thereby influencing electronic and magnetic properties. Given its success in equilibrium, strain can potentially provide a means to individually modify the response of coupled orders out of equilibrium.

\begin{figure}[!t]
    \centering
    \includegraphics[width=1\linewidth]{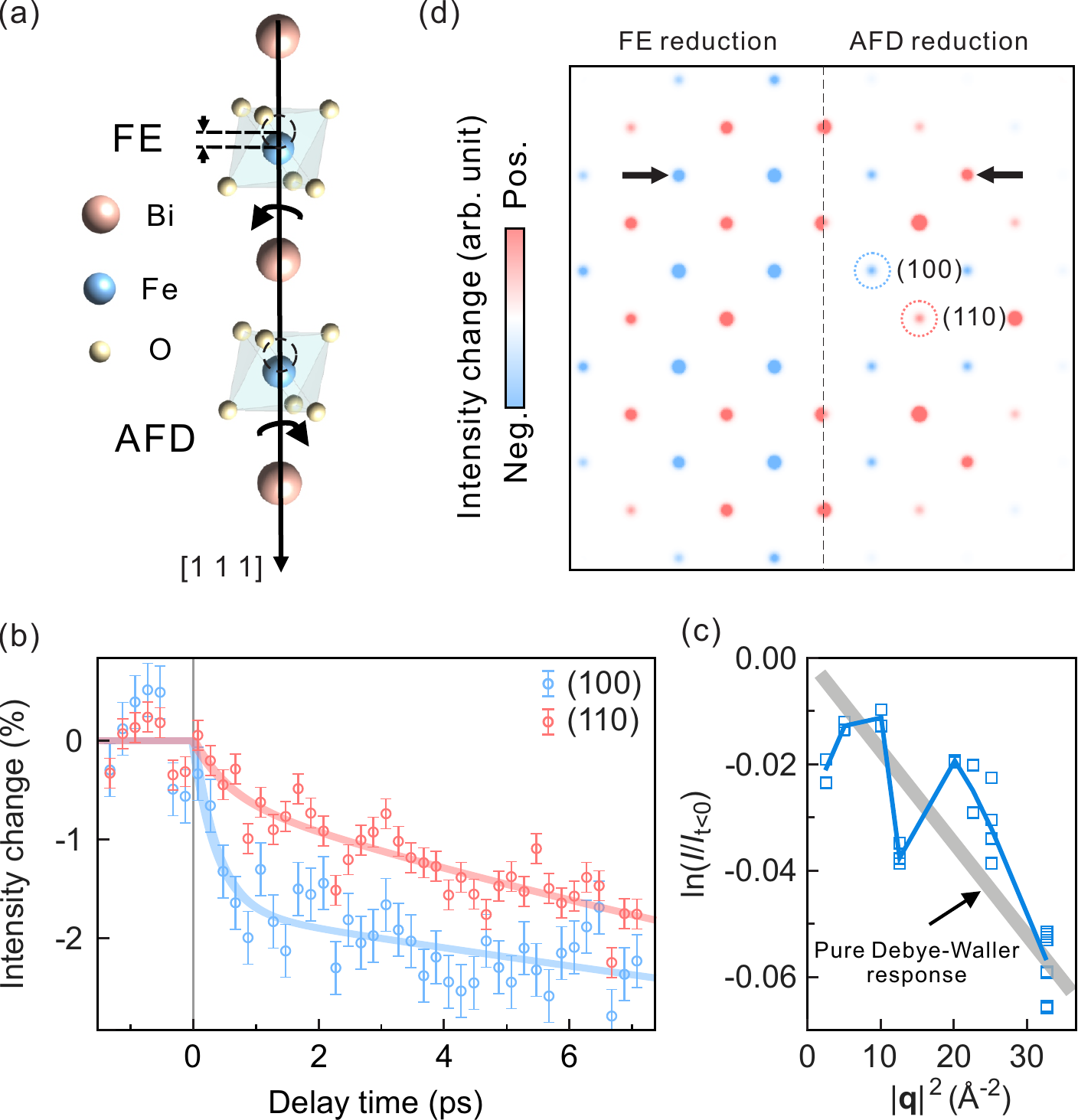}
    \caption{Two structural distortions associated with the multiferroic orders in BiFeO$_\text{3}$ and their effects on the diffraction pattern.
    (a)~Schematic of the off-center displacement of Fe atoms that underpins the ferroelectric (FE) order, and of the antiferrodistortive (AFD) rotation of oxygen octahedra that modulates the magnetic order.
    (b)~Photoinduced intensity changes of the (100) peak (blue) and the (110) peak (red) in the unstrained state after excitation by a 267-nm, 5-mJ/cm$^2$ pulse. Intensities are normalized by their averaged values before photoexcitation. Curves are fits to Eq.~\eqref{eq:S1} in the Supplemental Material. 
    (c)~Logarithmic intensity change $\ln[I(\mathbf{q},t)/I(\mathbf{q}, t<0)]$ for each Bragg peak, averaged over 5--7~ps and plotted as a function of $|\mathbf{q}|^2$ to assess whether the long-delay response is consistent with the Debye--Waller effect. {The blue line segments connect the average intensity change of peaks with the same $|\mathbf{q}|^2$, highlighting the deviation from the pure Debye--Waller response (gray line).}
    (d)~Simulated differential diffraction patterns when there is a reduction of the FE displacement~(\textit{left}) and the AFD rotation~(\textit{right}); the Debye--Waller effect is excluded in this simulation. Black arrows highlight the different effects by a reduction of FE displacement vs. AFD rotation.
    }
    \label{fig:sim_dif_pat}
\end{figure}

In this context, BiFeO$_3$ serves as an ideal system. It hosts ferroelectric order below $\sim 1,100~\mathrm{K}$ and antiferromagnetic order below $\sim 640~\mathrm{K}$~\cite{fischer1980temperature}. These multiferroic orders are highly sensitive to strain because they are both mediated by lattice distortions~\cite{zeches2009strain,Sando2014control_BFO,hemme2023tuning}. Specifically, the ferroelectric order is realized by a lattice mode with cation displacements that generate a large spontaneous polarization, known as the FE mode [Fig.~\ref{fig:sim_dif_pat}(a)]. On the other hand, due to Dzyaloshinskii--Moriya interactions, the magnetic order couples strongly to an antiferrodistortive (AFD) mode manifested by rotations of the oxygen octahedra [Fig.~\ref{fig:sim_dif_pat}(a)]~\cite{zhao2006electrical,chauleau2017multiSHGnm,liou2019deterministic}. Depending on the applied strain, these two lattice modes compete or cooperate via anharmonic interactions to determine the ground state of BiFeO$_3$ and, more generally, many other perovskite oxides~\cite{fechner2024quance_NMSTO,liou2019deterministic,Zhang2016cooper_NMLCMO,hemme2023tuning}.

To resolve the dynamics of the intertwined FE and AFD modes, we turned to MeV ultrafast electron diffraction (UED). This technique records a large number of Bragg peaks in a single image, enabling quantitative reconstruction of photoinduced lattice motions along the FE and AFD coordinates. However, UED requires freestanding membranes for transmission-geometry measurements, thereby precluding conventional strain-tuning approaches such as epitaxial growth. 

Here, we developed a robust transfer and \textit{in situ} straining method, despite the inherent mechanical fragility of ultrathin freestanding membranes. We successfully applied tunable, uniaxial strain to 30-nm-thick freestanding BiFeO$_3$ membranes with lateral sizes of more than $90~\upmu$m. Leveraging 44 Bragg peaks captured by MeV-UED, we quantitatively tracked the lattice dynamics associated with the FE and AFD distortions. We found that the FE displacement was suppressed on a sub-picosecond timescale, remarkably faster than the few-picosecond evolution of the AFD rotation. Crucially, tensile strain strongly reduced the photoinduced suppression of the FE displacement, whereas the nonequilibrium response of the AFD rotation remained strain-independent. These results demonstrate that strain can individually tune the ultrafast dynamics of otherwise coupled order parameters in multiferroics.

\begin{figure*}[!t]
    \centering
    \includegraphics[width=1\linewidth]{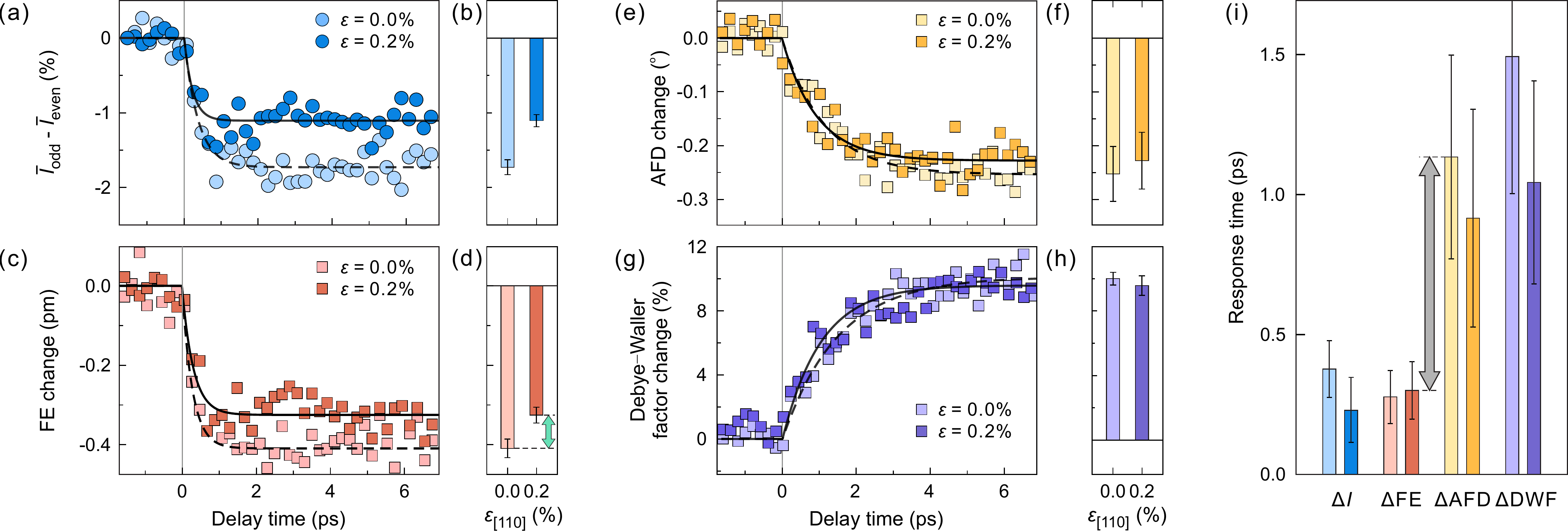}
    \caption{Lattice dynamics at different strain states. 
    (a,b) Photoinduced changes of (a)~the averaged differential intensities between $h+k$ = odd and even peaks and (b)~the long-time amplitudes. 
    (c--h)~Photoinduced structural changes extracted from the global fitting procedure (see \hyperref[sec:global_fitting]{Sec.~S3} in the Supplemental Material): (c)~FE displacement, (e)~AFD rotation angle, and (g)~Debye--Waller factor [DWF, as represented by $\langle u^2(t)\rangle$ in Eq.~\eqref{eq:DWF}], together with their corresponding long-time amplitudes (d,f,h) at different strain states. Markers represent experimentally extracted data while curves are fits to Eq.~\eqref{eq:S1} with only single-exponential dynamics (i.e., setting $A_2$ to zero). Light-colored (or dark-colored) markers and dashed (or solid) curves correspond to $\varepsilon_{[110]} = 0.0\%$ (or 0.2\%). (i)~Fitted response times for the different observables; $\Delta I \equiv \bar{I}_\text{odd} - \bar{I}_\text{even}$. The green double-sided arrow in (d) denotes the strain-induced difference in the transient FE suppression. The gray double-sided arrow in (i) denotes the difference in decay time constants between the FE displacement and AFD rotation. The long-time amplitudes in (b,d,f,h) are defined by parameter $C$ in Eq.~\eqref{eq:S1}; the response times in (i) reflect parameter $\tau_1$ in Eq.~\eqref{eq:S1}. Error bars in (b,d,f,h,i) are 1~s.d. of the fit parameters.
}
    \label{fig:dynamics_strain_scan}
\end{figure*}

We achieved continuously tunable strain on freestanding membranes with the following steps. To obtain the freestanding BiFeO$_3$, we first grew a BiFeO$_3$/Sr$_3$Al$_2$O$_6$/SrTiO$_3$ heterostructure via pulsed laser deposition [Fig.~\ref{fig:exp_setup}(a)], where the 10-nm-thick Sr$_3$Al$_2$O$_6$ was used as a sacrificial layer that was selectively etched in water~\cite{lu2016synthesisfreestandingNM}. After removing Sr$_3$Al$_2$O$_6$, the BiFeO$_3$ film was floated on the water surface and transferred onto a 90-$\upmu$m gap in a silicon chip mounted on a titanium plate, which has a large yield strain and a thermal expansion coefficient compatible with the strain cell (Razorbill Instruments). The titanium plate was then stressed by the piezo-actuated strain cell, transferring uniaxial strain to the suspended BiFeO$_3$ membrane along its [110] direction [Fig.~\ref{fig:exp_setup}(b,c)].

The MeV-UED setup with \textit{in situ} strain control is illustrated in Fig.~\ref{fig:exp_setup}(c). We first recorded static diffraction patterns while varying the strain-cell voltage, which controlled the applied strain. By tracking the shifts in Bragg peak positions, we directly determined the strain in the BiFeO$_3$ membrane based on the changes in the lattice parameters. Figure~\ref{fig:exp_setup}(d) shows the extracted strain $\varepsilon$ as a function of the voltage applied to the piezoelectric actuator. A maximum tensile strain of approximately 0.2\% was achieved; higher voltages were avoided to prevent membrane rupture. Despite the intrinsic hysteresis of the piezo-stacks, the recoverable strain curve demonstrates the robust performance of our strain device.

We varied the static strain between 0\% and 0.2\% as we photoexcited the sample with 267~nm (4.65~eV) laser pulses, where the photon energy exceeded the 2.7~eV bandgap of BiFeO$_3$~\cite{ihlefeld2008opticalgapBFO}. The probe consisted of electron pulses with 3.1~MeV kinetic energy. Figure~\ref{fig:sim_dif_pat}(b) shows the intensity evolutions of the (100) and (110) peaks following optical excitation without strain. Clear differences in decay timescales and amplitudes are evident: the (100) peak exhibits a faster and larger decrease, whereas the (110) peak shows a much slower and smaller reduction. This behavior is distinct from the expectation of thermal lattice heating, where the intensity dynamics would be governed by the Debye--Waller effect \cite{Cheng2022}, hence satisfying

\begin{equation}
\ln\left[\frac{I(\mathbf{q},t)}{I(\mathbf{q}, t<0)}\right] = -\frac{1}{3}|\mathbf{q}|^2\,\Delta\langle u^2(t)\rangle.
\label{eq:DWF}
\end{equation}
Here, $I(\mathbf{q},t)$ denotes the peak intensity at scattering vector $\mathbf{q}$ and delay time $t$, while $\Delta\langle u^2(t)\rangle$ represents the photoinduced change in the mean-square atomic displacement. If the Debye--Waller effect were the only response after photoexcitation, $\ln[I(\mathbf{q},t)/I(\mathbf{q}, t<0)]$ would scale linearly with $|\mathbf{q}|^2$ at any given delay time $t$ \cite{Cheng2022}, as indicated by the gray line in Fig.~\ref{fig:sim_dif_pat}(c). Instead, the measured $\ln[I(\mathbf{q},t)/I(\mathbf{q}, t<0)]$, shown as the blue squares and line segments in Fig.~\ref{fig:sim_dif_pat}(c), markedly deviates from this linear behavior, suggesting that the intensity changes cannot be solely accounted for by transient lattice heating, and additional structural changes must be considered.

To identify the transient structural modifications, we simulated diffraction pattern changes associated with reductions of the FE and AFD distortions, which are known to exhibit pronounced photoinduced dynamics in related oxide perovskites \cite{fechner2024quance_NMSTO}. As shown in Fig.~\ref{fig:sim_dif_pat}(d), in the absence of Debye--Waller heating, both FE and AFD distortions enhance the intensity of the (110) peak and suppress that of the (100) peak. These effects---when combined with the Debye--Waller intensity suppression---hence account for the smaller transient decrease of the (110) peak than the (100) peak, as experimentally observed in Fig.~\ref{fig:sim_dif_pat}(b).

A closer inspection of the FE simulation in the left panel of Fig.~\ref{fig:sim_dif_pat}(d) reveals that a reduction of the FE displacement induces opposite intensity changes in peaks with $h+k=$ even and odd. Hence, the difference between their averaged intensities, $\bar{I}_{\mathrm{odd}}-\bar{I}_{\mathrm{even}}$, provides a qualitative observable for the FE dynamics that can be directly extracted from the raw diffraction data. Here, $\bar{I}_{\mathrm{odd}}(t)\equiv I_\text{odd}(t) / I_\text{odd}(t<0)$, where $I_\text{odd}(t)$ is the integrated intensity of all visible Bragg peaks with $h+k=$ odd; $\bar{I}_{\mathrm{even}}(t)$ is similarly defined. Figure~\ref{fig:dynamics_strain_scan}(a) shows the time evolution of this observable for different initial strain states. We observed that the photoinduced evolution of $\overline{I}_{\text{odd}}-\overline{I}_{\text{even}}$ was smaller at $\varepsilon=0.2\%$ than at $\varepsilon=0.0\%$, suggesting that tensile strain suppresses the photoinduced reduction of the FE displacement.

To quantitatively separate the strain-dependent response of FE displacements, AFD rotations, and lattice heating, we performed a global fit using 44 Bragg peaks, as detailed in \hyperref[sec:crystal_modeling]{Secs.~S2--S3} in the Supplemental Material. The different momentum-space distributions of intensity changes associated with the FE displacement [Fig.~\ref{fig:sim_dif_pat}(d) \textit{left}], AFD rotation [Fig.~\ref{fig:sim_dif_pat}(d) \textit{right}], and the Debye--Waller effect [Fig.~\ref{fig:sim_dif_pat}(c)] provide the basis for disentangling their respective contributions. The extracted dynamics of each contribution are shown in Fig.~\ref{fig:dynamics_strain_scan}(c,e,g), reporting the photoinduced sub-picometer FE displacement and sub-degree AFD rotation, along with lattice heating represented by the percentage rise of the mean-square displacement.

In the absence of strain, the FE displacements respond with a time constant of 0.28(9)~ps, as shown in Figs.~\ref{fig:dynamics_strain_scan}(c,i). By contrast, the photoinduced dynamics of the AFD rotation is much slower, evolving with a time constant of 1.1(4)~ps [Figs.~\ref{fig:dynamics_strain_scan}(e,i)]. This separation of timescales is highlighted by the gray arrow in Fig.~\ref{fig:dynamics_strain_scan}(i), pointing to distinct mechanisms of the photoinduced change in the FE and AFD lattice distortions. Notably, the Debye--Waller factor increases on a comparable timescale to that of the AFD reduction [Figs.~\ref{fig:dynamics_strain_scan}(g,i)], suggesting that the AFD change is closely tied to lattice heating while the FE reduction has a nonthermal origin. 

Now we consider the case in which the strain was applied. The photoinduced reduction of the FE displacement is suppressed, as indicated by the green arrow in Fig.~\ref{fig:dynamics_strain_scan}(d). By contrast, the response of the AFD rotation shows no measurable strain dependence [Fig.~\ref{fig:dynamics_strain_scan}(f)], revealing distinct strain-modulated nonequilibrium responses of the FE and AFD distortions. Within experimental uncertainty, all decay time constants were found to be independent of strain [Fig.~\ref{fig:dynamics_strain_scan}(i)]. By analyzing the peak positions as a function of time delay, we further found that the photoinduced change in strain is negligible regardless of the initial strain state (Fig.~\ref{fig:strainvstimedelay}). These observations indicate that the initial strain state prior to photoexcitation can alter the nonequilibrium dynamics of the FE displacements without changing that of the AFD rotations.

\begin{figure*}[!t]
    \centering
    \includegraphics[width=1\linewidth]{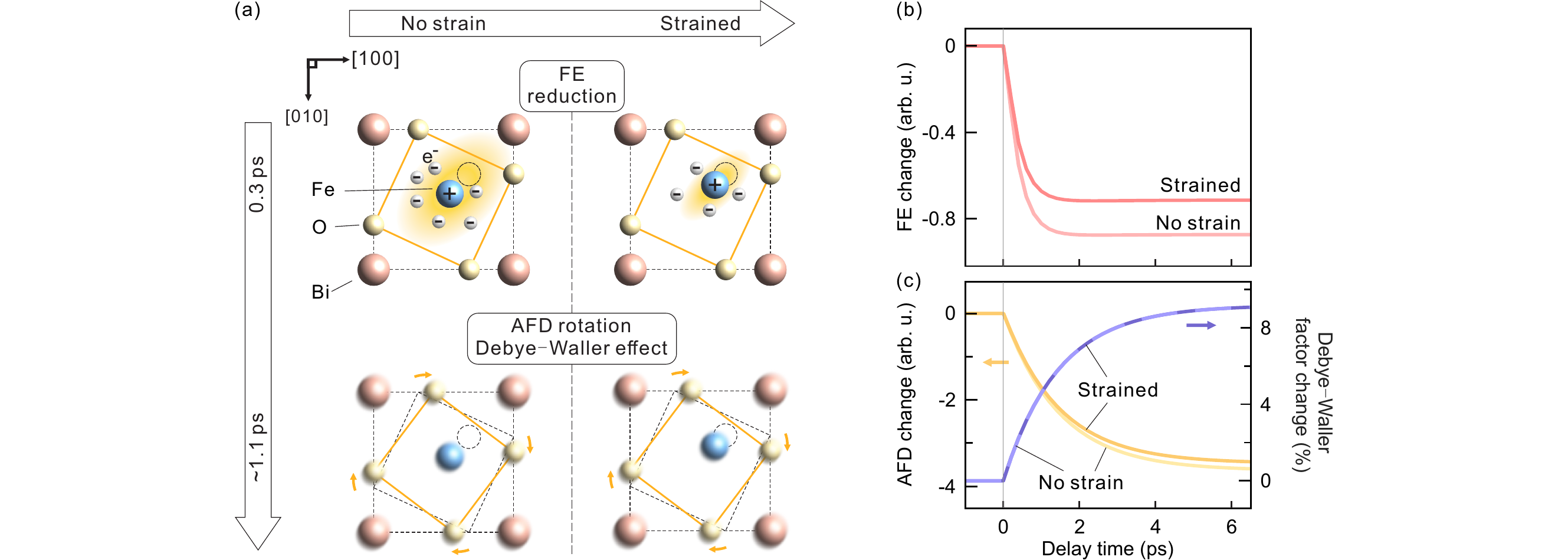}
    \caption{Distinct mechanisms for photoinduced dynamics of ferroelectric displacement and antiferrodistortive rotation. 
    (a)~Schematic illustration of the strain-selective atomic pathways following above-gap photoexcitation, shown without strain (\textit{left} column) and with tensile strain (\textit{right} column). Dashed circles indicate the initial Fe positions, orange arrows denote the oxygen octahedral rotation in the AFD mode, and blurred atoms represent isotropic thermal vibrations. In the top right panel, the strain-induced reduction of mobile electron density (yellow shade) signifies weaker carrier screening, which selectively affects the ferroelectric displacement but not the AFD rotation.
    (b,c)~Simulated temporal evolutions of (b)~FE displacement, 
    (c)~AFD octahedral rotation (left axis), and the Debye--Waller factor (right axis). Simulations were based on the time-dependent Ginzburg--Landau model described in \hyperref[sec:tdgl_model]{Sec.~S5} in the Supplemental Material.
    }
    \label{fig:LG_sim}
\end{figure*}

To understand this distinction, we note that FE displacements generate electric polarizations that can be screened by photoexcited mobile carriers, during which the lattice remains relatively cold. The characteristic timescale for this screening effect to occur has been found within 0.2~ps in BiFeO$_3$ \cite{Chen2025SmallpolaronPhRvX_XUV,paillard2016photostriction_PRLBFO}. The fast generation of mobile carriers is followed by their energy transfer to the lattice, as reflected in the Debye–Waller lattice heating on the 1--1.5~ps timescale [Fig.~\ref{fig:dynamics_strain_scan}(g,h,i)]. The excited carriers remain mobile until electrons and holes recombine across the indirect bandgap, a process that can take over 100~ns \cite{Yamada2014PRB_BFOphotocarriers}. Hence, the polar FE displacement remains screened over an extended period after photoexcitation, and the FE suppression persists well beyond 70~ps in our long-delay measurement (Fig.~\ref{fig:longtimetrace}).

Unlike the polar FE displacement, the AFD rotation generates no net electric polarization and does not respond directly to photoinduced screening, leading instead to slower dynamics on the timescale of lattice heating as reflected in the photoinduced Debye--Waller effect [Fig.~\ref{fig:dynamics_strain_scan}(g,h,i)]. The schematic structural evolution is illustrated in Fig.~\ref{fig:LG_sim}(a). These distinct mechanisms underlying the photoinduced dynamics of the FE and AFD distortions point to a strain-tunable screening effect that selectively modifies the transient FE displacement while leaving AFD rotations intact.

We developed a time-dependent Ginzburg--Landau model to provide a phenomenological understanding of the strain-modulated FE and AFD responses (see \hyperref[sec:tdgl_model]{Sec.~S5} in the Supplemental Material for details). By numerically solving the equations of motion, we found that the experimentally observed strain dependence of the FE suppression and the lack of such dependence in the AFD dynamics were reproduced \textit{only} when the photoinduced screening effect was hindered by tensile strain. Importantly, we tested alternative scenarios in which the photoinduced screening term was kept strain-independent while other Landau parameters, including the FE and AFD anharmonic coefficients and the FE–AFD coupling coefficient, were allowed to vary freely with strain. As demonstrated in Fig.~\ref{fig:LGfixedAlpha}, these alternative simulations all fail to reproduce the key experimental trend that strain selectively modifies the FE displacement but not the AFD rotation. The simulated dynamics with and without strain are shown in Figs.~\ref{fig:LG_sim}(b) and (c). In agreement with the experiments, the strained case exhibits a reduced suppression of the FE displacement, while the AFD reduction remains unchanged under strain. Our analysis further shows that the AFD dynamics is instead dominated by the Debye--Waller effect (Fig.~\ref{fig:AFDcontribution}), and both exhibit a similar timescale in their ultrafast responses [Fig.~\ref{fig:LG_sim}(c)].

One plausible physical origin of the strain-dependent screening effect underlying the FE dynamics is transient small-polaron formation and associated carrier trapping. The generation of such polarons has been demonstrated in BiFeO$_3$ by previous transient extreme-ultraviolet spectroscopy and time-resolved sum-frequency generation measurements, which further show that polaron formation enhances carrier self-trapping and weakens the screening of the ferroelectric polarization~\cite{Chen2025SmallpolaronPhRvX_XUV}. As reported in a similar ferroelectric perovskite, BaTiO$_3$, strain can promote polaron formation by lowering the carrier self-trapping energy \cite{Xu2019BTO_strain_polaronnpjCM}. Consequently, tensile lattice strain can trap mobile carriers into polarons, weakening the photoinduced screening and hence reducing the suppression of the FE order [Fig.~\ref{fig:LG_sim}(a)].

In summary, we demonstrated that \textit{in situ} strain can selectively tune the nonequilibrium multiferroic dynamics in freestanding BiFeO$_3$ membranes. This capability is made possible by the simultaneous capture of the FE and AFD dynamics in the same detector image, which unambiguously shows that tensile strain selectively weakens the photoinduced suppression of the FE displacement but leaves the AFD response intact. This contrasting strain dependence is rooted in the distinct driving mechanisms: the FE dynamics is caused by carrier screening, while the AFD dynamics is dominated by transient lattice heating. Our results show that coupled lattice distortions in a multiferroic can be individually steered in a nonequilibrium regime by tuning strain. More broadly, adding independent control knobs effectively promotes dynamical trajectories to a higher-dimensional phase space, offering a promising strategy for optimizing nonequilibrium transition pathways.

\vspace{0.23cm}
\noindent\textit{Acknowledgments} --- We thank Yu-Che Chien and Jacob Ruiz for sample characterization. We thank Jiarui Li, Yonghun Lee, Zhengyan Darius Shi, Antonio Picano, Daniel Teitelman, and Joonho Lee for helpful discussions. We thank Ying Chen and Aaron Garza for integrating the strain cell into the cryogenic environment. We also thank Joel England, Sharon S. Philip, Fuhao Ji, and Stephen P. Weathersby for their support during the MeV-UED beamtime. This work is primarily supported by the U.S. Department of Energy, Office of Basic Energy Sciences under award No.~DE-SC0026202. H.Y.H. and M.H. acknowledge funding from the U.S. Department of Energy, Office of Basic Energy Sciences, Division of Materials Sciences and Engineering, under contract No.~DE-AC02-76SF00515. The MeV-UED facility is operated as part of the Linac Coherent Light Source at the SLAC National Accelerator Laboratory, supported by the U.S. Department of Energy, Office of Science, Office of Basic Energy Sciences under contract No.~DE-AC02-76SF00515.

\vspace{0.23cm}
\noindent\textit{Data availability} --- The data supporting the findings of this study are included in the main text and Supplemental Material and are also openly available in Dryad \cite{BFOstrain_data}. Further information is available from the corresponding author upon reasonable request.

\clearpage

\onecolumngrid

\setcounter{secnumdepth}{2}
\setcounter{section}{0}
\setcounter{subsection}{0}
\setcounter{figure}{0}
\setcounter{table}{0}
\setcounter{equation}{0}
\renewcommand{\theequation}{S\arabic{equation}}
\renewcommand{\thefigure}{S\arabic{figure}}
\renewcommand{\thetable}{S\arabic{table}}
\renewcommand{\theHequation}{S\arabic{equation}}
\renewcommand{\theHfigure}{S\arabic{figure}}
\renewcommand{\theHtable}{S\arabic{table}}
\renewcommand{\thesection}{S\arabic{section}}
\renewcommand{\thesubsection}{S\arabic{section}.\arabic{subsection}}

\begin{center}
{\large\bfseries Supplemental Material for}\\[0.5em]
{\large\bfseries ``Selective strain tuning of nonequilibrium multiferroic dynamics in BiFeO$_\text{3}$''}\\[0.5em]
\end{center}

\vspace{0.5em}

\section{S\MakeLowercase{ample preparation and experimental setup}}
\label{sec:sample_setup}
Single-crystal BiFeO$_3$ films were epitaxially grown on the (001) surface of SrTiO$_3$ (STO) substrates by pulsed laser deposition (PLD) using a KrF excimer laser ($\lambda = 248$~nm, Coherent Inc.) and a polycrystalline ceramic BiFeO$_3$ target. The base pressure of the PLD chamber was $5 \times 10^{-10}$~Torr. Prior to deposition, the STO(001) substrates were annealed at $1,000^{\circ}\mathrm{C}$ for 1~hour in $1 \times 10^{-6}$~Torr of oxygen to obtain a clean and atomically flat surface. A 10-nm-thick Sr$_3$Al$_2$O$_6$ (SAO) sacrificial layer was then deposited at $880^{\circ}\mathrm{C}$ under an oxygen pressure of $5 \times 10^{-6}$~Torr, with the thickness calibrated by reflection high-energy electron diffraction oscillations. During the growth, the laser repetition rate was 5~Hz and the fluence was 1.5~J/cm$^2$. Subsequently, a 30-nm-thick BiFeO$_3$ film was deposited on the SAO layer at $850^{\circ}\mathrm{C}$ in 10~mTorr of oxygen, using a laser repetition rate of 5~Hz and a fluence of 1~J/cm$^2$. After growth, the SAO layer was dissolved in deionized water to release the BiFeO$_3$ film, which was then transferred onto a silicon chip with a gap, mounted on a titanium plate for strain application.

In our UED measurement, the electron beam with 3.1~MeV kinetic energy was produced in a photocathode radio-frequency gun operating at 360~Hz. 267-nm laser pulses ($\sim$65~fs, full width at half maximum, FWHM) were used for photoexcitation. The diameter of the pump laser spot was $\sim$300~$\upmu$m (FWHM). The pump laser was at normal incidence to the sample, parallel to the electron probe pulses. The size of the electron probe pulse at the sample location was approximately 100~$\upmu$m, and the pulse width was approximately 150~fs (FWHM). Further details of the UED beamline can be found in Ref.~\cite{WEATH2015megaSLACbeamline}.

\section{C\MakeLowercase{rystal Structure Modeling}}
\label{sec:crystal_modeling}
We constructed a crystal structure model of BiFeO$_3$ to simulate the diffraction patterns. A $2\times2\times2$ pseudocubic supercell was used, with the coordinate origin defined at the Bi site. The lattice constants are $a=b=c=3.947$~\AA~\cite{pal2010BiFeO3}. To describe the structure of BiFeO$_3$ concisely, we adopted the Glazer notation for oxygen octahedral rotations. In this notation, octahedral rotations are specified by three projected angles about the $a$, $b$, and $c$ axes. For the \textit{R}3c structure of BiFeO$_3$, these three angles are equal; we therefore denote the rotation angles by a single angle $\theta$. We further define $\delta_{\mathrm{Fe}}$ as the Fe-site displacement along the polarization direction and $\delta_{\mathrm{O}}$ as the corresponding displacement of the oxygen octahedra. 

The atomic configurations of one particular ferroelectric domain are summarized in Table~\ref{tab:BFO-structure}, and other domains are generated using the symmetry operations. Because the transverse coherence length of MeV electrons is only a few nanometers~\cite{Zhu2015femto,park2022NanoL..22.9275P}, which is substantially smaller than the ferroelectric domain size~\cite{physi2009AdM....21.2463C}, electron scattering from different domains adds incoherently. We therefore compute the squared structure factor for each domain independently and then average over all domains. In different domains, the magnitudes of Fe displacements and the angles of oxygen octahedral rotations are assumed to be identical. To simplify the fitting model, we fixed the ratio $\delta_\textrm{Fe}/\delta_\textrm{O}$, which is collectively described by an FE displacement, such that BiFeO$_3$ reaches the paraelectric phase when $\delta_\textrm{Fe} = 0$.

\begin{figure*}[b!]
    \centering
    \includegraphics[width=0.72\linewidth]{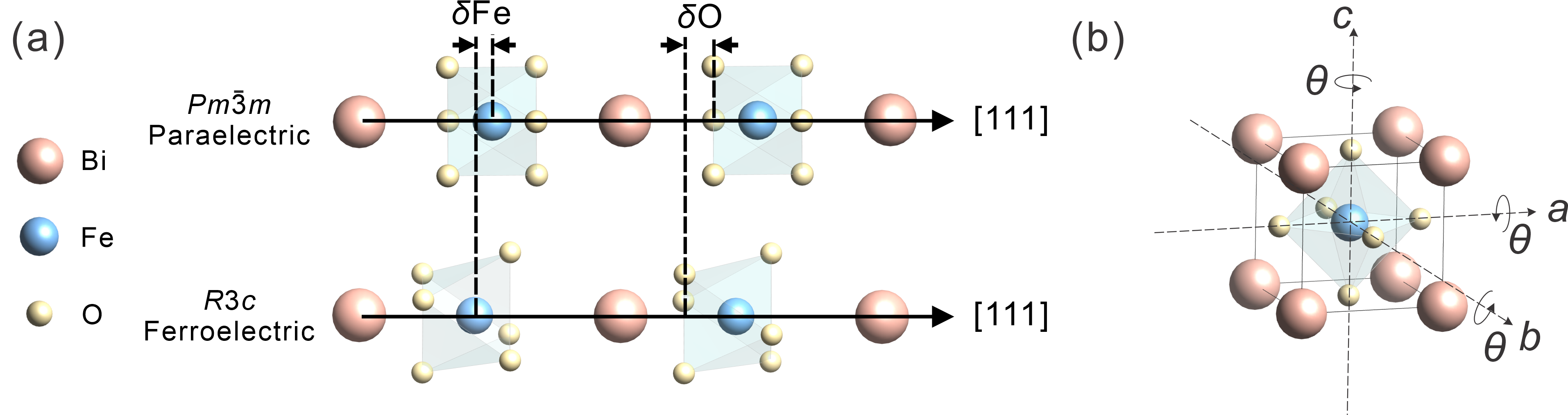}
    \caption{(a)~Crystal structure of BiFeO$_3$ in the paraelectric (top) and ferroelectric (bottom) phases.  (b)~Glazer notation description of the antiferrodistortive oxygen octahedral rotations in BiFeO$_3$, which are tied to the magnetism in this material.
}\label{fig:BFOstructure}
\end{figure*}

\begin{ThreePartTable}
\begin{longtable}[p]{cccc}
\caption{Parameterized $2\times2\times2$ pseudocubic supercell crystal structure of BiFeO$_3$ in fractional coordinates. These coordinates correspond to the [111] domain of the ferroelectric order.} 
\label{tab:BFO-structure} 
\\  \toprule
    \toprule
    Atom    &   $x$  & $y$   & $z$      \\ 
    \midrule
    \endfirsthead
    \toprule
    \toprule
    Atom    &   {x}  & {y}   & {z}      \\ 
    \midrule
    \endhead
    \hline
    \hline
    \endfoot
    \endlastfoot
    Bi1        &0   &0   &0  \\ 
    Bi2        &0.5 &0   &0  \\ 
    Bi3        &0   &0.5 &0  \\
    Bi4        &0.5 &0.5 &0  \\
    Bi5        &0   &0   &0.5\\ 
    Bi6        &0.5 &0   &0.5\\
    Bi7        &0   &0.5 &0.5\\
    Bi8        &0.5 &0.5 &0.5\\
    Fe1        &0.25 + $\delta_{\mathrm{Fe}}$ &0.25 + $\delta_{\mathrm{Fe}}$ &0.25 + $\delta_{\mathrm{Fe}}$\\
    Fe2        &0.25 + $\delta_{\mathrm{Fe}}$ &0.75 + $\delta_{\mathrm{Fe}}$ &0.25 + $\delta_{\mathrm{Fe}}$\\ 
    Fe3        &0.75 + $\delta_{\mathrm{Fe}}$ &0.25 + $\delta_{\mathrm{Fe}}$ &0.25 + $\delta_{\mathrm{Fe}}$\\
    Fe4        &0.75 + $\delta_{\mathrm{Fe}}$ &0.75 + $\delta_{\mathrm{Fe}}$ &0.25 + $\delta_{\mathrm{Fe}}$\\
    Fe5        &0.25 + $\delta_{\mathrm{Fe}}$ &0.25 + $\delta_{\mathrm{Fe}}$ &0.75 + $\delta_{\mathrm{Fe}}$\\ 
    Fe6        &0.25 + $\delta_{\mathrm{Fe}}$ &0.75 + $\delta_{\mathrm{Fe}}$ &0.75 + $\delta_{\mathrm{Fe}}$\\ 
    Fe7        &0.75 + $\delta_{\mathrm{Fe}}$ &0.25 + $\delta_{\mathrm{Fe}}$ &0.75 + $\delta_{\mathrm{Fe}}$\\ 
    Fe8        &0.75 + $\delta_{\mathrm{Fe}}$ &0.75 + $\delta_{\mathrm{Fe}}$ &0.75 + $\delta_{\mathrm{Fe}}$\\
    O1 &0.25 -- 0.25$\rm{sin}\theta$ + $\delta_{\mathrm{O}}$ & 0.25 + 0.25$\mathrm{sin}\theta$ + $\delta_{\mathrm{O}}$ & 0 + $\delta_{\mathrm{O}}$                     \\ 
    O2 &0.75 + 0.25$\mathrm{sin}\theta$ + $\delta_{\mathrm{O}}$ & 0.25 -- 0.25$\mathrm{sin}\theta$ + $\delta_{\mathrm{O}}$ & 0 + $\delta_{\mathrm{O}}$                      \\ 
    O3 &0.25 + 0.25$\mathrm{sin}\theta$ + $\delta_{\mathrm{O}}$ & 0.75 -- 0.25$\mathrm{sin}\theta$ + $\delta_{\mathrm{O}}$ & 0 + $\delta_{\mathrm{O}}$                      \\
    O4 &0.75 -- 0.25$\mathrm{sin}\theta$ + $\delta_{\mathrm{O}}$ & 0.75 + 0.25$\mathrm{sin}\theta$ + $\delta_{\mathrm{O}}$ & 0 + $\delta_{\mathrm{O}}$                      \\
    O5 &0 + $\delta_{\mathrm{O}}$                       & 0.25 -- 0.25$\mathrm{sin}\theta$ + $\delta_{\mathrm{O}}$ & 0.25 + 0.25$\mathrm{sin}\theta$ + $\delta_{\mathrm{O}}$\\ 
    O6 &0 + $\delta_{\mathrm{O}}$                       & 0.75 + 0.25$\mathrm{sin}\theta$ + $\delta_{\mathrm{O}}$ & 0.25 -- 0.25$\mathrm{sin}\theta$ + $\delta_{\mathrm{O}}$\\
    O7 &0.25 + 0.25$\mathrm{sin}\theta$ + $\delta_{\mathrm{O}}$ & 0 + $\delta_{\mathrm{O}}$                       &0.25 -- 0.25$\mathrm{sin}\theta$ + $\delta_{\mathrm{O}}$   \\
    O8 &0.25 -- 0.25$\mathrm{sin}\theta$ + $\delta_{\mathrm{O}}$ & 0.5 + $\delta_{\mathrm{O}}$                     &0.25 + 0.25$\mathrm{sin}\theta$ + $\delta_{\mathrm{O}}$   \\
    O9        &0.5 + $\delta_{\mathrm{O}}$                     &0.25 + 0.25$\mathrm{sin}\theta$  + $\delta_{\mathrm{O}}$&0.25 -- 0.25$\mathrm{sin}\theta$ + $\delta_{\mathrm{O}}$\\
    O10        &0.5 + $\delta_{\mathrm{O}}$                     &0.75 -- 0.25$\mathrm{sin}\theta$  + $\delta_{\mathrm{O}}$&0.25 + 0.25$\mathrm{sin}\theta$ + $\delta_{\mathrm{O}}$\\ 
    O11        &0.75 -- 0.25$\mathrm{sin}\theta$ + $\delta_{\mathrm{O}}$ &0                        + $\delta_{\mathrm{O}}$&0.25 + 0.25$\mathrm{sin}\theta$ + $\delta_{\mathrm{O}}$\\
    O12        &0.75 + 0.25$\mathrm{sin}\theta$ + $\delta_{\mathrm{O}}$ &0.5                      + $\delta_{\mathrm{O}}$&0.25 -- 0.25$\mathrm{sin}\theta$ + $\delta_{\mathrm{O}}$\\
    O13        &0.25 + 0.25$\mathrm{sin}\theta$ + $\delta_{\mathrm{O}}$ &0.25 -- 0.25$\mathrm{sin}\theta$  + $\delta_{\mathrm{O}}$&0.5 + $\delta_{\mathrm{O}}$\\ 
    O14        &0.75 -- 0.25$\mathrm{sin}\theta$ + $\delta_{\mathrm{O}}$ &0.25 + 0.25$\mathrm{sin}\theta$  + $\delta_{\mathrm{O}}$&0.5 + $\delta_{\mathrm{O}}$\\ 
    O15        &0.25 -- 0.25$\mathrm{sin}\theta$ + $\delta_{\mathrm{O}}$ &0.75 + 0.25$\mathrm{sin}\theta$  + $\delta_{\mathrm{O}}$&0.5 + $\delta_{\mathrm{O}}$\\ 
    O16        &0.75 + 0.25$\mathrm{sin}\theta$ + $\delta_{\mathrm{O}}$ &0.75 -- 0.25$\mathrm{sin}\theta$  + $\delta_{\mathrm{O}}$&0.5 + $\delta_{\mathrm{O}}$\\
    O17        &0 + $\delta_{\mathrm{O}}$                       &0.25 + 0.25$\mathrm{sin}\theta$  + $\delta_{\mathrm{O}}$&0.75 -- 0.25$\mathrm{sin}\theta$ + $\delta_{\mathrm{O}}$\\ 
    O18        &0    + $\delta_{\mathrm{O}}$                    &0.75 -- 0.25$\mathrm{sin}\theta$  + $\delta_{\mathrm{O}}$&0.75 + 0.25$\mathrm{sin}\theta$ + $\delta_{\mathrm{O}}$\\
    O19        &0.25 -- 0.25$\mathrm{sin}\theta$ + $\delta_{\mathrm{O}}$ &0                        + $\delta_{\mathrm{O}}$&0.75 + 0.25$\mathrm{sin}\theta$ + $\delta_{\mathrm{O}}$\\
    O20        &0.25 + 0.25$\mathrm{sin}\theta$ + $\delta_{\mathrm{O}}$ &0.5                      + $\delta_{\mathrm{O}}$&0.75 -- 0.25$\mathrm{sin}\theta$ + $\delta_{\mathrm{O}}$\\
    O21        &0.5  + $\delta_{\mathrm{O}}$                    &0.25 -- 0.25$\mathrm{sin}\theta$  + $\delta_{\mathrm{O}}$&0.75 + 0.25$\mathrm{sin}\theta$ + $\delta_{\mathrm{O}}$\\ 
    O22        &0.5 + $\delta_{\mathrm{O}}$                     &0.75 + 0.25$\mathrm{sin}\theta$  + $\delta_{\mathrm{O}}$&0.75 -- 0.25$\mathrm{sin}\theta$ + $\delta_{\mathrm{O}}$\\
    O23        &0.75 + 0.25$\mathrm{sin}\theta$ + $\delta_{\mathrm{O}}$ &0                        + $\delta_{\mathrm{O}}$&0.75 -- 0.25$\mathrm{sin}\theta$ + $\delta_{\mathrm{O}}$\\
    O24        &0.75 -- 0.25$\mathrm{sin}\theta$ + $\delta_{\mathrm{O}}$ &0.5                      + $\delta_{\mathrm{O}}$&0.75 + 0.25$\mathrm{sin}\theta$ + $\delta_{\mathrm{O}}$\\
    \bottomrule
    \bottomrule
    \end{longtable}
\end{ThreePartTable}

Under an electron beam incident along [001], certain domain variants become indistinguishable because they share identical in-plane polarization projections. Moreover, antiparallel polarization domains yield identical diffraction patterns. Consequently, the four variants $[111]$, $[11\bar{1}]$, $[\bar{1}\bar{1}1]$, and $[\bar{1}\bar{1}\bar{1}]$ produce identical diffraction patterns; the same applies to $[1\bar{1}1]$, $[1\bar{1}\bar{1}]$, $[\bar{1}11]$, and $[\bar{1}1\bar{1}]$. Therefore, only two distinct domain types are considered in the simulations, denoted as $[111]$-type and $[1\bar{1}1]$-type domains.

\section{R\lowercase{etrieval of transient atomic trajectories via a global fitting procedure}}
\label{sec:global_fitting}
The intensity of the Bragg peaks for a single domain can be calculated using the kinematic scattering theory~\cite{kirkland1998advanced}:

\begin{equation}
\begin{aligned}
    I &\propto |F|^2 =\left|\sum_j 
         \exp\left(- \frac{|\mathbf{q}|^2 \left \langle u_j^2 \right \rangle}{6} \right) f_j  \exp\left[-i2\pi\left(h x_j + k y_j + l z_j\right)\right]\right|^2,
\end{aligned}
\label{eq3}
\end{equation}
where $F$ is the structure factor and the summation over $j$ runs through all atoms in the unit cell (8~Bi atoms, 8~Fe atoms, and 24~O atoms). The first exponential term in Eq.~\eqref{eq3} is the Debye--Waller factor, where $\left \langle u_j^2 \right \rangle$ is the isotropic mean-squared displacement for the $j$th atom associated with lattice heating. The relative amplitudes for different atomic elements are fixed according to neutron-scattering measurements at room temperature, with the ratio ${ \langle u^2_{\mathrm{Bi}}  \rangle :  \langle u^2_{\mathrm{Fe}}  \rangle :  \langle u^2_{\mathrm{O}}  \rangle} = 0.53 : 0.41 : 0.71$~\cite{pal2010BiFeO3}. In Eq.~\eqref{eq3}, $|\mathbf{q}|$ denotes the magnitude of the momentum transfer of a peak, $f_j$ is the atomic scattering factor for the $j$th atom, $\left(x_j, y_j, z_j\right)$ are the fractional coordinates of the $j$th atom in the unit cell, and $\left(hkl\right)$ are the Miller indices of the diffraction peak. From Eq.~\eqref{eq3}, we can obtain 44 equations from the intensities of 44 Bragg peaks in each diffraction pattern, where the experimental diffraction intensities are given by $I_{\mathrm{exp}} = \eta I_{[111]} + (1- \eta)I_{[1\bar{1}1]}$. Here $I_{[111]}$ and $I_{[1\bar{1}1]}$ denote the diffraction intensities of $[111]$- and $[1\bar{1}1]$-type domains, respectively, and $\eta$ denotes the domain fraction. At each time delay, we use the least-squares method to minimize the residual $R$, determining the best solution for the structural parameters $\delta_{\mathrm{Fe}}$, $\delta_{\mathrm{O}}$, $\theta$, and $\left \langle u_j^2 \right \rangle$ that we used to generate the structure of BiFeO$_3$. Here, \textit{R} is defined as:

\begin{equation}
   R = \frac{\sum_m \left\{ \mathrm{log}(I^{\textrm{exp}}_m) - \mathrm{log}(I^{\textrm{sim}}_m)\right\}^2}{\sum_m \left\{\mathrm{log}(I^{\textrm{exp}}_m)\right\}^2},
\end{equation}
where the summation runs over all 44 Bragg peaks in each diffraction pattern, $I^{\textrm{exp}}_m$ is the measured intensity of the $m$th peak, and $I^{\textrm{sim}}_m$ is the simulated intensity of the $m$th peak.

\begin{figure}[t!]
    \centering
    \includegraphics[width=0.98\linewidth]{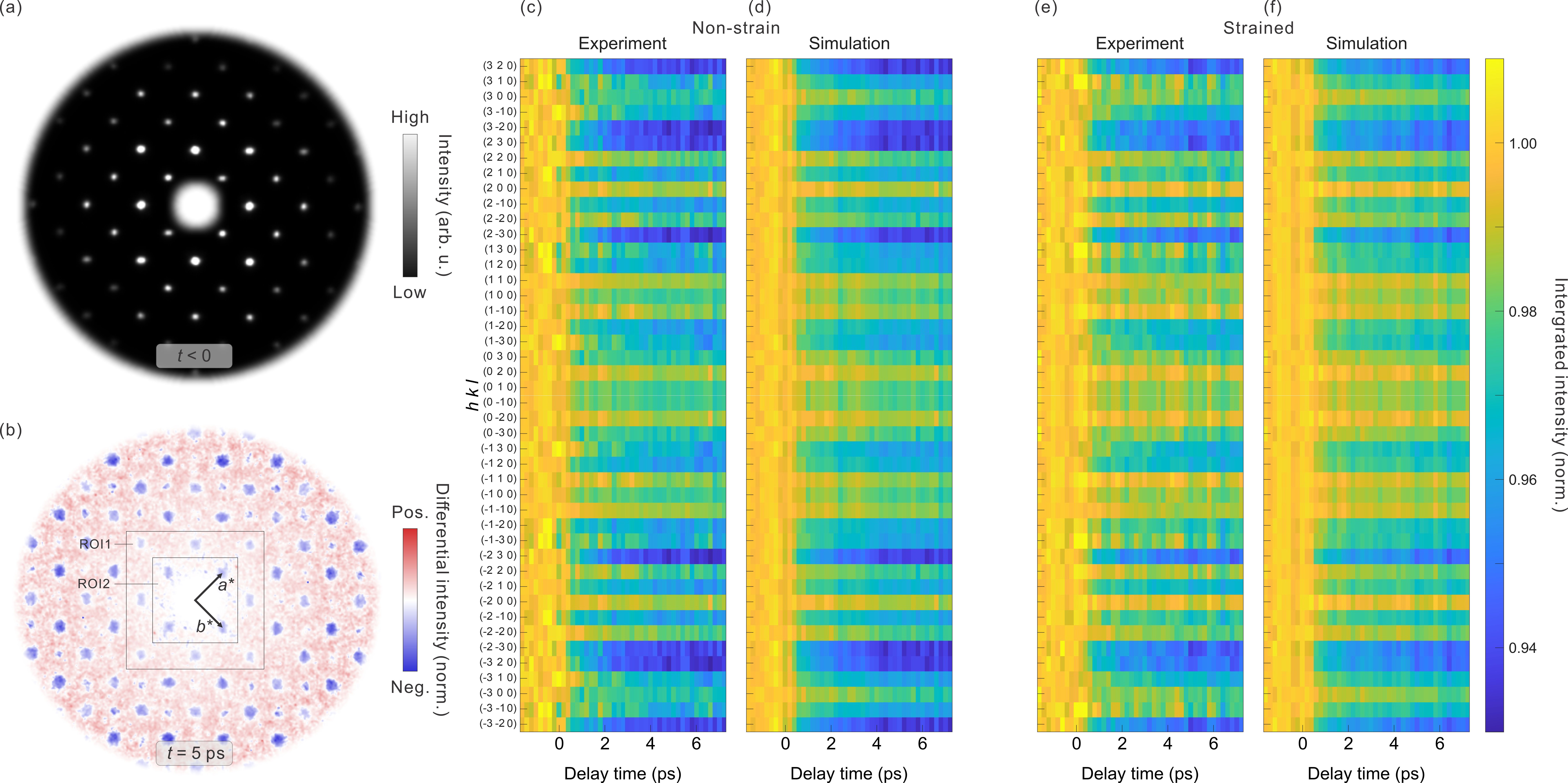}
    \caption{{~(a)~Diffraction pattern of BiFeO$_3$ before the arrival of the pump pulse ($t<0$). (b)~Differential diffraction pattern at 5~ps pump-probe delay. The sample in both (a) and (b) was in the unstrained state. In (b), ROI1 includes the group of \{200\} and \{110\} peaks with even $h+k$, whereas ROI2 includes the group of \{100\} peaks with odd $h+k$. It is worth noting that the peak intensities in ROI1 decrease less than those in ROI2. In addition, in the regions outside ROI1 and ROI2, the intensity suppression alternates between smaller and larger changes. (c--f)~Direct comparison between (c,~e)~experimental and (d,~f)~simulated peak intensities calculated using the fitting parameters from Fig.~\ref{fig:dynamics_strain_scan}. The Miller indices $(hkl)$ of each peak are included in (c). The data in (c,~d) are for the unstrained sample, while those in (e,~f) are for the sample under a 0.2\% tensile strain. The intensity values are normalized by their respective averaged values before the arrival of the pump laser pulse.}}
    \label{fig:fitresultcomp}
\end{figure}

\begin{figure}
    \centering
    \includegraphics[width=0.35\linewidth]{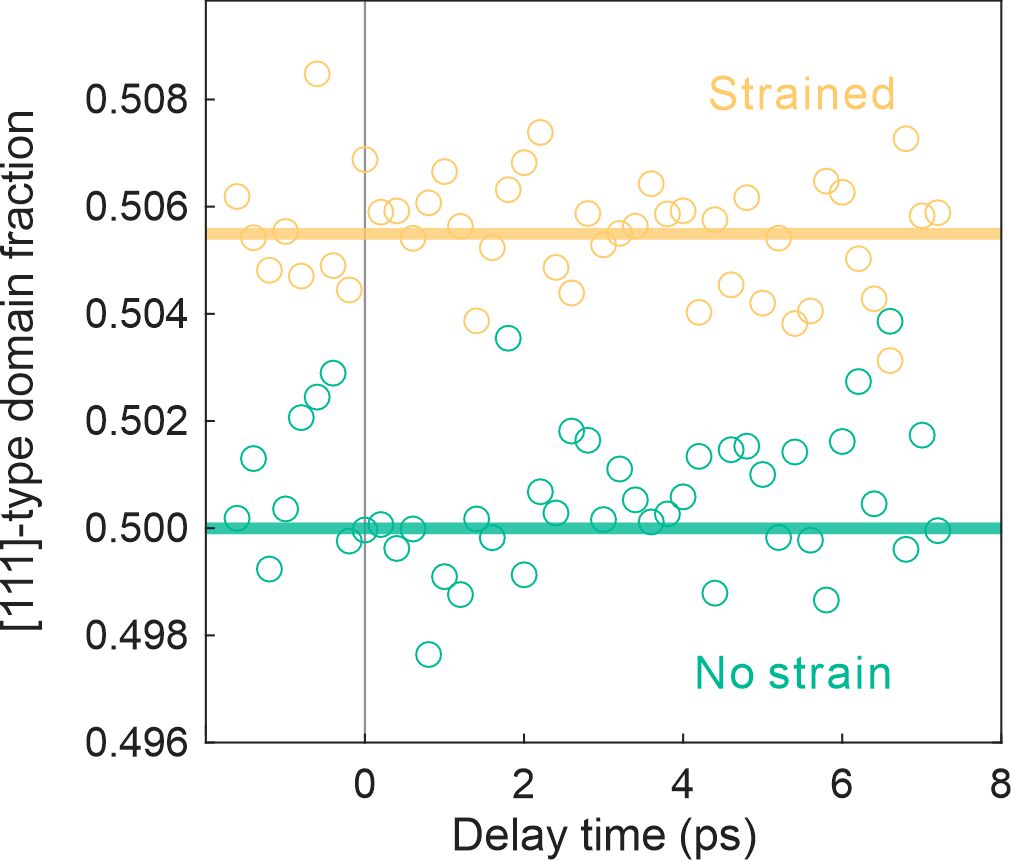}
    \caption{Extracted fraction of the [111]-type domains with~(yellow) and without~(green) the initial strain. Before strain is applied, the two domain types have nearly equal populations, as expected for a freestanding membrane without external constraints. When strain is applied along the [110] direction, the fraction of the [111]-type domains becomes larger than that of the [1$\bar{1}$1]-type domains. This increase is consistent with the previous report that tensile strain tends to align the polarization with the strain direction~\cite{peng2020SciAdv}. The domain fractions are time-independent within experimental uncertainty.
}\label{fig:Domainfrac.jpg}
\end{figure}

The global fitting results are shown in Fig.~\ref{fig:fitresultcomp}(d,f) and Fig.~\ref{fig:dynamics_strain_scan}. Before strain is applied, the two domain types have nearly equal populations, as expected for a freestanding membrane without external constraints (Fig.~\ref{fig:Domainfrac.jpg}). When strain is applied along the [110] direction, the fraction of the [111]-type domain becomes larger than that of the $[1\bar{1}1]$-type domain, as expected from the alignment of the polarization with the strain direction \cite{peng2020SciAdv}. The domain fractions are time-\emph{independent} within experimental uncertainty (Fig.~\ref{fig:Domainfrac.jpg}). As shown in Fig.~\ref{fig:fitresultcomp}(c--f), there is good agreement between experimental peak intensities and simulated peak intensities calculated using the fitted parameters from Fig.~\ref{fig:dynamics_strain_scan}.

\section{P\lowercase{henomenological model for analyzing time traces}}
\label{sec:phenomenological_model}
The transient signal is modeled using a step-function onset followed by two exponential response channels. The intensity evolution is written as

\begin{equation}
I(t)=H(t-t_0)\left[
A_1 e^{-\frac{t-t_0}{\tau_1}}
+ A_2 e^{-\frac{t-t_0}{\tau_2}}
+ C
\right],
\label{eq:S1}
\end{equation}
where $H(t)$ is the Heaviside step function describing the arrival of the pump pulse at time $t_0$, $A_{1,2}$ and $\tau_{1,2}$ are the amplitudes and characteristic response times of a fast and a slow component, respectively, and $C$ represents a long-lived offset. The curves in Fig.~\ref{fig:sim_dif_pat}(b) are obtained when both $A_1$ and $A_2$ are nonzero to capture both the fast FE component and the slow AFD/Debye--Waller component. The curves in Figs.~\ref{fig:dynamics_strain_scan}(a,c,e,g) are obtained by keeping $A_2$ at zero to extract the characteristic response times in the different observables.

\section{T\lowercase{ime-dependent \MakeUppercase{G}inzburg--\MakeUppercase{L}andau model}}
\label{sec:tdgl_model}
The coupled dynamics of the FE and AFD order parameters are described by the inertial time-dependent Ginzburg--Landau equations of motion:

\begin{align}
m_1 \ddot{\psi}_1 + \gamma_1 \dot{\psi}_1 + 
\frac{\partial f}{\partial \psi_1} &= 0, 
\\[4pt]
m_2 \ddot{\psi}_2 + \gamma_2 \dot{\psi}_2 + 
\frac{\partial f}{\partial \psi_2} &= 0,
\end{align}
where $\psi_1(t)$ and $\psi_2(t)$ denote the FE and AFD order parameters, respectively. Here $m_{1,2}$ are the effective inertial masses and $\gamma_{1,2}$ are phenomenological damping coefficients. The total Landau free-energy density is taken as \cite{fedorova2022PRB_phasefield_BFO,cheng2020defects} 

\begin{equation}
\begin{aligned}
f(\psi_1,\psi_2,\varepsilon) &=
\underbrace{\tfrac{1}{2}\alpha_1 \psi_1^2 + \tfrac{1}{4}\beta_1 \psi_1^4 }_{\text{FE contribution}}
+ \underbrace{\tfrac{1}{2}\alpha_2 \psi_2^2 + \tfrac{1}{4}\beta_2 \psi_2^4 }_{\text{AFD contribution}} \\[4pt]
&\quad + \underbrace{g_{12}\, \psi_1^2 \psi_2^2}_{\text{FE--AFD coupling}}
+ \underbrace{(g_{1\varepsilon} \psi_1^2 + g_{2\varepsilon} \psi_2^2)\,\varepsilon}_{\text{electro- and rotostrictive coupling}}\\
\\
&= \tfrac{1}{2}\alpha_{1,\mathrm{eff}} \psi_1^2
+ \tfrac{1}{4}\beta_1 \psi_1^4
+ \tfrac{1}{2}\alpha_{2,\mathrm{eff}} \psi_2^2
+ \tfrac{1}{4}\beta_2 \psi_2^4
+ g_{12}\, \psi_1^2 \psi_2^2 .
\end{aligned}
\label{eq:F_total}
\end{equation}
Here $\alpha_{1,\mathrm{eff}}\equiv\alpha_1+2g_{1\varepsilon}\varepsilon$ and $\alpha_{2,\mathrm{eff}}\equiv\alpha_2+2g_{2\varepsilon}\varepsilon$. $\alpha_{1,\mathrm{eff}}$ and $\alpha_{2,\mathrm{eff}}$ are time-dependent quadratic coefficients; $\beta_{1,2}$ are time-independent anharmonic coefficients. The parameter $g_{12}$ describes the biquadratic coupling between the two order parameters, while $g_{1\varepsilon}$ and $g_{2\varepsilon}$ quantify their coupling to strain $\varepsilon$. 

\begin{figure*}[!b]
    \centering
    \includegraphics[width=0.6\linewidth]{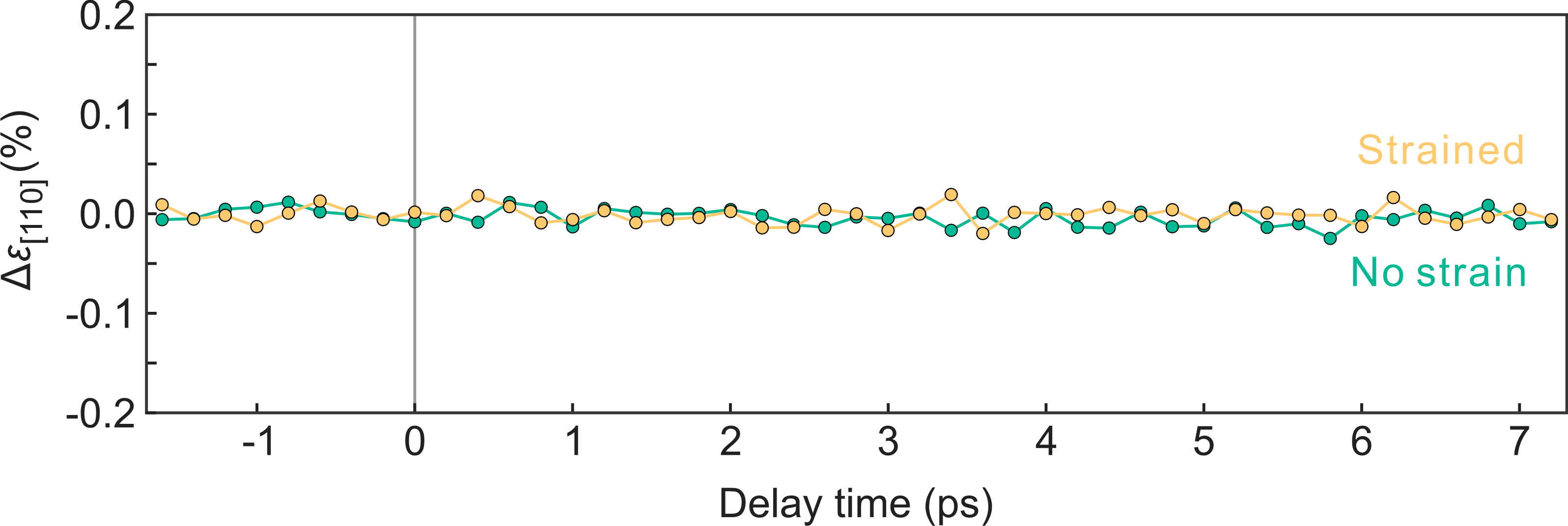}
    \caption{Evolution of the photoinduced change in the [110] strain, calculated based on the Bragg peak positions. Within our experimental sensitivity, the strain remains the same before and after photoexcitation in the probed window.
}
    \label{fig:strainvstimedelay}
\end{figure*}

In the time-dependent case, the coefficients $\alpha_{1,\mathrm{eff}}$ and $\alpha_{2,\mathrm{eff}}$ evolve in time following photoexcitation and can be decomposed into contributions with distinct timescales. Prior to excitation ($t<0$), both coefficients remain constant at $\alpha_{1,\mathrm{eff},t<0}$ and $\alpha_{2,\mathrm{eff},t<0}$, respectively. After the pump ($t\ge0$), $\alpha_{1,\mathrm{eff}}(t)$ and $\alpha_{2,\mathrm{eff}}(t)$ are modified by adding fast and slow components to these pre-pump values:

\begin{align}
\Delta\alpha_{1,\mathrm{eff}}(t) &= H(t)\!\left[\,\Delta\alpha_{1,\mathrm{eff}}^{\mathrm{screen}}
+\Delta\alpha_{1,\mathrm{eff}}^{\mathrm{thermal}}\!\left(1 - e^{-t/\tau_{\mathrm{thermal}}}\right) \right],
\label{eq:dalpha1} \\
\Delta\alpha_{2,\mathrm{eff}}(t) &= H(t)\!\left[\Delta\alpha_{2,\mathrm{eff}}^{\mathrm{thermal}}\!\left(1 - e^{-t/\tau_{\mathrm{thermal}}}\right)\right].
\label{eq:dalpha2}
\end{align}
Here, $\Delta\alpha_{1,\mathrm{eff}}^{\mathrm{screen}}$ is a constant that represents an instantaneous shift of the equilibrium FE displacement due to carrier screening. This screening process is expected to occur on a sub-200-fs timescale~\cite{Chen2025SmallpolaronPhRvX_XUV,paillard2016photostriction_PRLBFO}, which is shorter than our instrumental response; we hence treat it as instantaneous in our model. $\Delta\alpha_{1,\mathrm{eff}}^{\mathrm{thermal}}$ and $\Delta\alpha_{2,\mathrm{eff}}^{\mathrm{thermal}}$ describe slower thermal renormalizations with characteristic time constant $\tau_{\mathrm{thermal}}$. $\Delta\alpha_{2,\mathrm{eff}}(t)$ does not include a fast screening component because the AFD rotation does not respond directly to the photoinduced screening, but instead evolves on a timescale governed by the Debye--Waller effect. $\Delta\alpha_{2,\mathrm{eff}}^{\mathrm{thermal}}$ is strain-independent because the observed Debye--Waller effect is identical with and without strain. The photoinduced strain as a function of time is measured by the peak shifts, revealing that $\varepsilon$ is time-independent as shown in Fig.~\ref{fig:strainvstimedelay}; we thus treat $\varepsilon$ as a time-independent input of the model. Under these conditions, the complete parameter set provides a self-consistent description of the coupled FE--AFD potential landscape, dissipation, and pump-induced perturbations under different strain states.

The dynamics of $\psi_1$ and $\psi_2$ are determined by the effective driving forces:

\begin{align}
\frac{\partial f}{\partial \psi_1}
&=\left[\alpha_{1,\mathrm{eff,t<0}}+\Delta\alpha_{1,\mathrm{eff}}(t)\right]
\psi_1
+\beta_1\psi_1^3
+2g_{12}\psi_1\psi_2^2,
\label{eq:df_dpsi1}\\
\frac{\partial f}{\partial \psi_2}
&=\left[\alpha_{2,\mathrm{eff,t<0}}+\Delta\alpha_{2,\mathrm{eff}}(t)\right]\!\psi_2
+\beta_2\psi_2^3
+2g_{12}\psi_2\psi_1^2.
\label{eq:df_dpsi2}
\end{align}

\begin{figure*}[!t]
    \centering
    \includegraphics[width=0.7\linewidth]{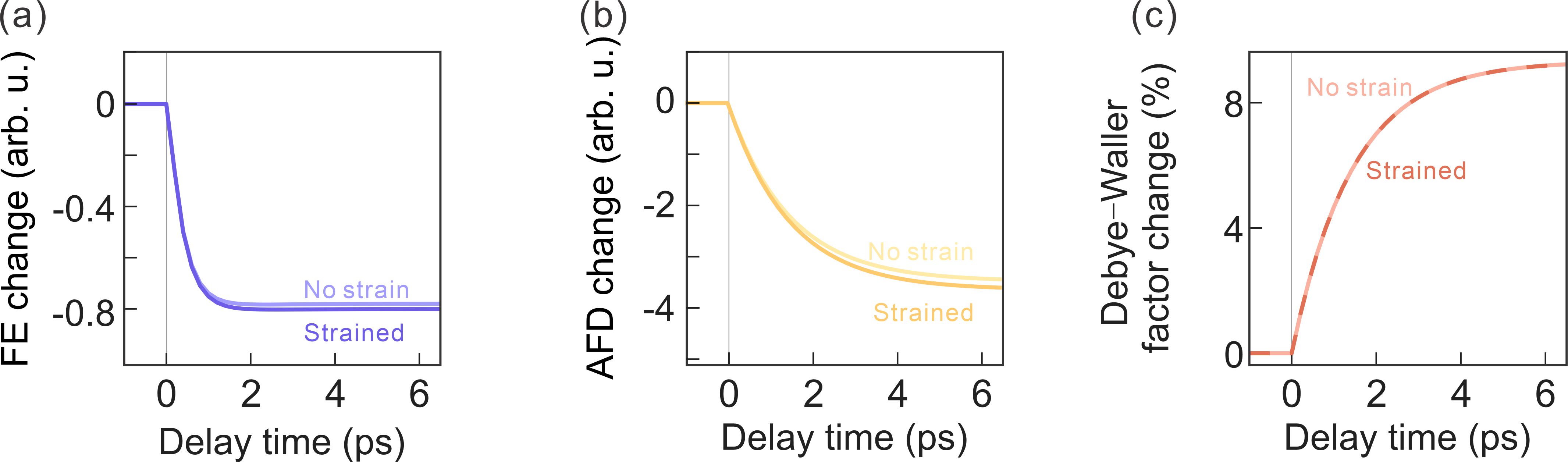}
    \caption{Illustration of the failure to fit the experimental data in Fig.~\ref{fig:dynamics_strain_scan}(c,e,g) for the dynamics of (a)~FE displacements, (b)~AFD rotations, and (c)~the Debye--Waller effect (DWF). These fits are obtained when $\Delta\alpha_{1,\mathrm{eff}}^{\mathrm{screen}}$ is fixed to be strain-independent while anharmonic coefficients $\beta_1$ and $\beta_2$ and the FE--AFD coupling coefficient $g_{12}$ are allowed to vary freely with strain. The failure to reproduce the strain-dependent FE response while keeping the AFD response strain-independent suggests that $\Delta\alpha_{1,\mathrm{eff}}^{\mathrm{screen}}$ is necessarily modulated by strain.
}
    \label{fig:LGfixedAlpha}
\end{figure*}

Numerical solutions to the model show that reproducing the experimental dynamics requires a strain-induced reduction of $\Delta\alpha_{1,\mathrm{eff}}^{\mathrm{screen}}$. If we instead fix $\Delta\alpha_{1,\mathrm{eff}}^{\mathrm{screen}}$ to be strain-independent, the experimentally observed trends cannot be reproduced by the model, even when $\beta_1$, $\beta_2$, and $g_{12}$ are allowed to vary freely. As an illustration, the dynamics simulated using fixed $\Delta\alpha_{1,\mathrm{eff}}^{\mathrm{screen}}$ and freely varying $\beta_1$, $\beta_2$, and $g_{12}$ are shown in Fig.~\ref{fig:LGfixedAlpha}. This set of trends fails to reproduce the strain-induced suppression of the FE reduction while keeping the AFD response nearly unchanged.

\begin{figure*}[!h]
    \centering
    \includegraphics[width=0.3\linewidth]{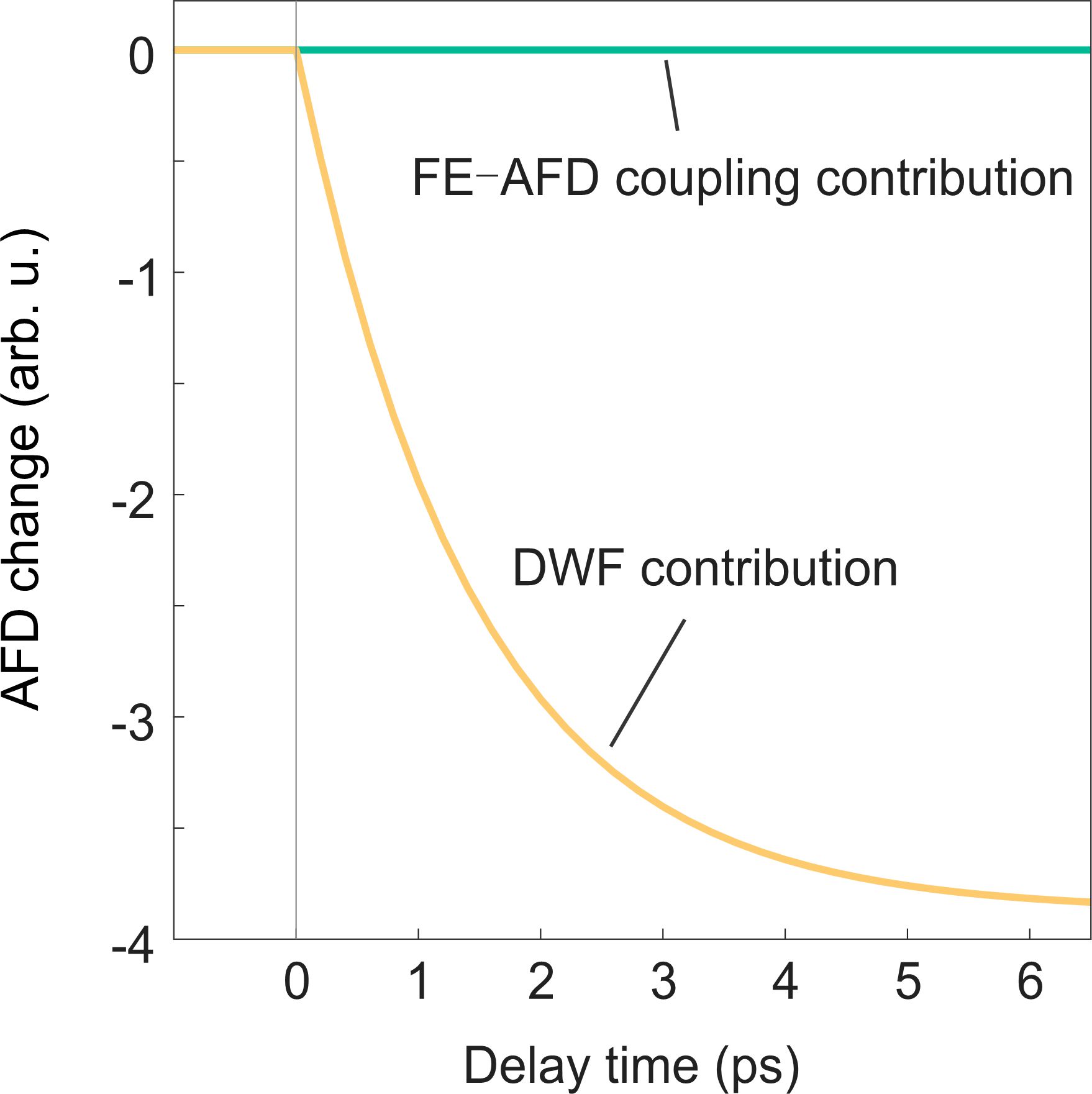}
    \caption{Comparison of the different contributions to the transient AFD rotation. Based on our time-dependent Ginzburg--Landau model, it can be driven by both the anharmonic FE--AFD coupling term and the Debye--Waller effect in Eq.~\eqref{eq:df_dpsi2}. We manually set $\Delta\alpha_{2,\mathrm{eff}}(t) = 0$ to get the FE--AFD coupling component (green curve), and set $g_{12}=0$ to get the Debye--Waller contribution (yellow curve). The substantial change of the yellow curve compared with the green curve suggests that the Debye--Waller effect plays a dominant role in the AFD dynamics. The curves are plotted for the unstrained case; the same conclusion was reached for the strained film using this analysis.
}
    \label{fig:AFDcontribution}
\end{figure*}

Using the fitting parameters that reproduce the experimental observations, based on Eqs.~\eqref{eq:dalpha2} and \eqref{eq:df_dpsi2}, we can further decompose the AFD rotation dynamics into contributions from the anharmonic FE--AFD coupling ($2g_{12}\psi_1^2\psi_2$) and Debye--Waller heating ($\Delta\alpha_{2,\mathrm{eff}}(t)\psi_2$). To isolate the FE--AFD coupling contribution, we first suppose that the AFD order parameter $\psi_2$ is driven solely by the fast dynamics of $\psi_1$ through the term $2g_{12}\psi_1^2\psi_2$ by setting $\Delta\alpha_{2,\mathrm{eff}}(t)=0$. The resulting dynamics are shown as the green curve in Fig.~\ref{fig:AFDcontribution}. On the other hand, by setting $g_{12}=0$, the evolution of $\psi_2$ arises only from the Debye--Waller contribution, shown as the yellow curve in Fig.~\ref{fig:AFDcontribution}. The amplitude reduction induced by the Debye--Waller heating is significantly larger than that by FE--AFD coupling. From this analysis, our experimental observation is consistent with the picture that the Debye--Waller effect plays a dominant role in the transient AFD dynamics.

\section{S\MakeLowercase{train-induced \MakeUppercase{FE} and \MakeUppercase{AFD} changes in the equilibrium state}}
\label{sec:equilibrium_strain}

\begin{figure*}[!t]
     \centering
    \includegraphics[width=0.7\linewidth]{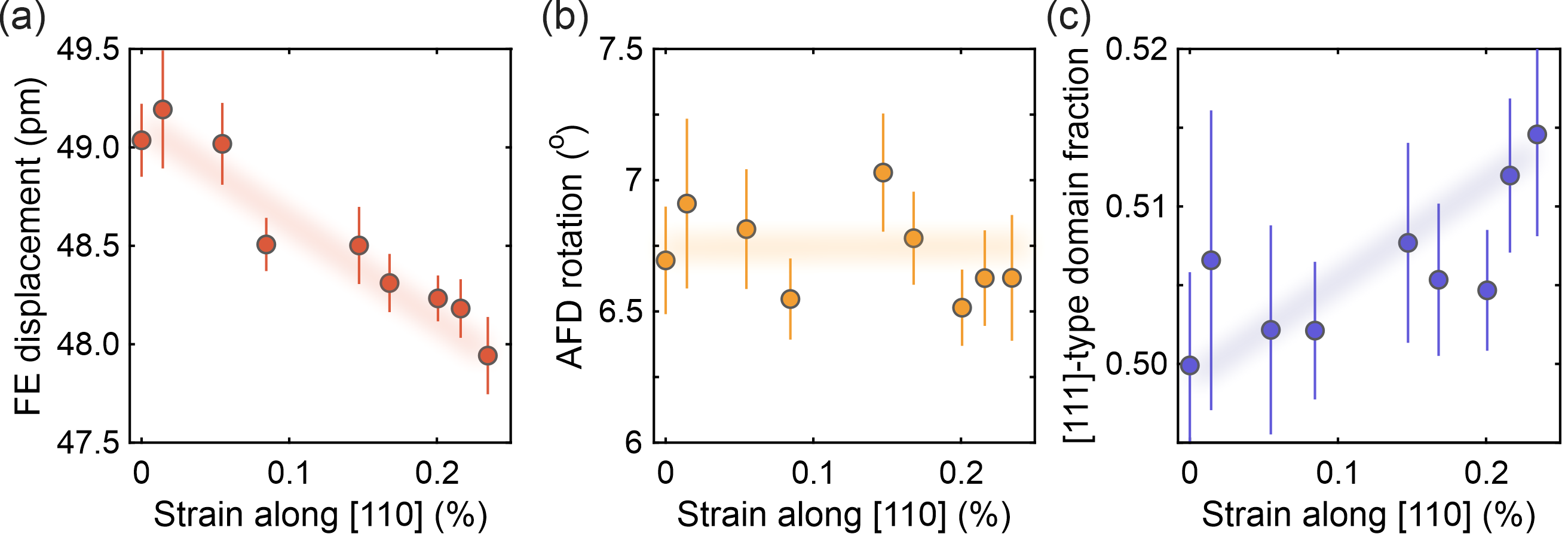}
    \caption{Static strain-induced changes to the equilibrium (a) FE displacement, (b) AFD rotation angle, and (c) the fraction of [111]-type domains. These quantities are extracted based on the global fitting procedures described in \hyperref[sec:global_fitting]{Sec.~S3}. Error bars correspond to 1~s.d. of the global fitting uncertainty; shaded lines are guides to the eye.}
    \label{fig:strainscanFE1}
\end{figure*}

In this section, we study the effect of tensile strains on the static structure prior to photoexcitation. In this analysis, a diffraction pattern was recorded at each strain level using the same dataset as that plotted in Fig.~\ref{fig:exp_setup}(d), and the global fitting analysis described in \hyperref[sec:global_fitting]{Sec.~S3} was carried out to extract the FE displacement, AFD rotation, and domain fractions. To improve the signal-to-noise ratio, we binned data points within $\pm 0.1$~V of the strain cell voltage and averaged the ramp-up and -down scans. The results are summarized in Fig.~\ref{fig:strainscanFE1}: there is a systematic strain-induced reduction of the FE displacement, whereas the AFD rotation remains nearly unchanged within the noise level. In addition, there is a slight increase in the [111]-type domain fraction, consistent with the previous report that tensile strain tends to align the polarization with the strain direction \cite{peng2020SciAdv}.

To clarify the microscopic origin of the strain-dependent FE response in equilibrium, we consider strain-induced local structural inhomogeneity in freestanding membranes. In a freestanding membrane, applied strain is often accommodated by the formation of dislocation-induced defects, which broadens the distribution of local structural distortions \cite{Schlom2014ElasticStrain}. Ferroelectricity involves a collective structural mode stabilized by the coherent long-range dipolar interaction among local polar displacements. Random local strain fields, defect potentials, and domain-wall pinning can disrupt this coherence and reduce the electrostatic energy gain associated with a uniformly polarized state. As a result, the local FE potential can be renormalized toward a shallower double-well potential, leading to a smaller local polar displacement. By contrast, the AFD rotation is a nonpolar oxygen-octahedral mode and is therefore not directly affected by the disruption of long-range polar coherence caused by local structural inhomogeneity. Thus, within our experimental strain range, AFD rotation exhibits a much weaker strain dependence prior to photoexcitation.

The above distinction between polar vs. nonpolar nature of the FE displacement vs. AFD rotation not only accounts for their different responses to static strain in equilibrium but also underlies their distinct photoinduced behavior as a function of static strain, as we elaborated in the main text. In particular, while the polar FE displacement sensitively responds to a changing Coulomb environment after laser excitation due to photoinduced carrier screening, the nonpolar AFD rotation does not. Since static strain-induced local disorder and defects can facilitate carrier self-trapping and small-polaron formation \cite{Emin1994PRB_disorder_polaron}, they can effectively reduce photoinduced carrier screening and hence suppress the transient reduction of the FE displacement. On the other hand, the nonpolar AFD rotation is not as sensitive to strain-mediated carrier screening as the polar FE displacement, and its photoinduced dynamics hence remains largely intact as a function of static strain, as shown in Fig.~\ref{fig:dynamics_strain_scan}(e,f).

\begin{figure*}[!b]
     \centering
    \includegraphics[width=1\linewidth]{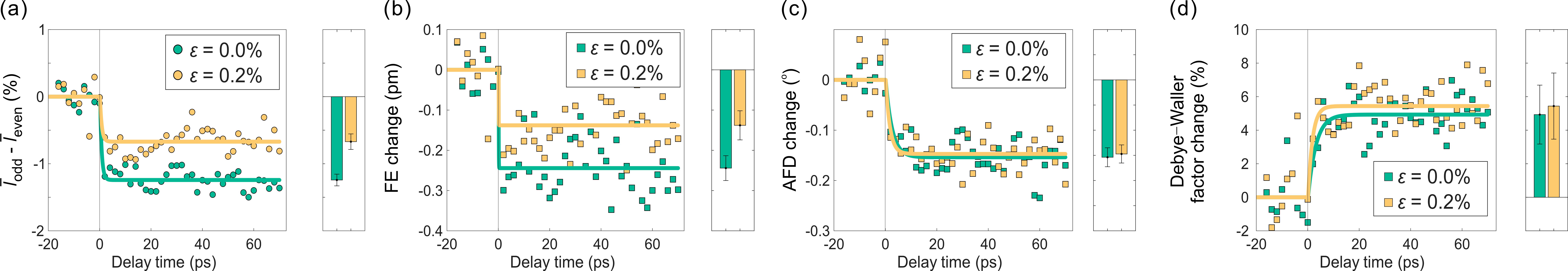}
    \caption{Long-time evolutions of (a)~$\bar{I}_{\mathrm{odd}}-\bar{I}_{\mathrm{even}}$, (b)~FE displacement change, (c)~AFD rotation change, and (d)~the Debye--Waller factor change measured at a lower pump fluence of $3~\mathrm{mJ/cm^2}$ compared to the data shown in Fig.~\ref{fig:dynamics_strain_scan}. Green and yellow symbols represent data measured at $\varepsilon = 0.0\%$ and $\varepsilon = 0.2\%$, respectively. Solid curves are fits to the experimental data using Eq.~\eqref{eq:S1} with only single-exponential dynamics. The bar plots on the right of each panel summarize the fitted amplitudes. Under tensile strain, the $\bar{I}_{\mathrm{odd}}-\bar{I}_{\mathrm{even}}$ contrast and FE response are suppressed, whereas the AFD rotation and Debye--Waller factor show no measurable strain dependence, reproducing the selective strain-mediated dynamics observed in Fig.~\ref{fig:dynamics_strain_scan}.
}
\label{fig:longtimetrace}
\end{figure*}

\vspace{1.0cm}
\section{S\MakeLowercase{train-selective nonequilibrium lattice dynamics at low pump fluence and long time delay}}
\label{sec:low_fluence_long_delay}

All data shown in the main text were taken with 5~mJ/cm$^2$ incident pump laser fluence for a pump--probe delay window up to 7~ps. To verify the strain-selective behavior of the transient multiferroic dynamics, here we show the results taken at a lower pump fluence of $3~\mathrm{mJ/cm^2}$ over a more extended time window up to $70~\mathrm{ps}$. Similar to the observations in Fig.~\ref{fig:dynamics_strain_scan}, upon applying tensile strain, both photoinduced $\bar{I}_{\mathrm{odd}}-\bar{I}_{\mathrm{even}}$ [Fig.~\ref{fig:longtimetrace}(a)] and the FE displacement change [Fig.~\ref{fig:longtimetrace}(b)] are suppressed, whereas the AFD rotation [Fig.~\ref{fig:longtimetrace}(c)] and the Debye--Waller factor [Fig.~\ref{fig:longtimetrace}(d)] show no measurable strain dependence. 

Compared with the data measured at $5~\mathrm{mJ/cm^2}$ in Fig.~\ref{fig:dynamics_strain_scan}, all observables in Fig.~\ref{fig:longtimetrace} exhibit smaller photoinduced changes at $3~\mathrm{mJ/cm^2}$. The reduced FE response is consistent with weaker carrier-induced screening due to the smaller number of mobile carriers generated at lower pump fluence. Similarly, the Debye--Waller factor increases by approximately $6\%$, smaller than the $\sim 10\%$ increase observed at $5~\mathrm{mJ/cm^2}$ in Fig.~\ref{fig:dynamics_strain_scan}(g), consistent with reduced laser-induced lattice heating. Although the dataset shown in Fig.~\ref{fig:longtimetrace} has a lower signal-to-noise ratio due to the lower pump fluence, it reproduces the same key observation as the one shown in the main text: tensile strain selectively suppresses the transient FE response while leaving the ultrafast dynamics of AFD rotation and Debye–Waller factor nearly unchanged.
\end{document}